\documentclass[referee]{aa} % for a referee version
\usepackage{graphicx,aalongtable}
\usepackage{txfonts}
\usepackage{natbib}
\bibpunct{(}{)}{;}{a}{}{,}
\begin{document}
   \title{High resolution spectroscopy of stars with transiting planets\thanks{Based 
   on observations collected at the ESO 8.2-m VLT-UT2 Kueyen telescope (program ID\,75.C-0185).}}

   \subtitle{The cases of OGLE-TR-10, 56, 111, 113, and TrES-1}

   \author{N.C.~Santos\inst{1,2} \and
	   F.~Pont\inst{2} \and 
           C.~Melo\inst{3} \and
	   G.~Israelian\inst{4} \and 
	   F.~Bouchy\inst{5,6} \and 
	   M.~Mayor\inst{2} \and
	   C.~Moutou\inst{5} \and
	   D.~Queloz\inst{2} \and 
	   S.~Udry\inst{2} \and
	   T.~Guillot\inst{7}
           }

   \offprints{N.C. Santos, \email{nuno@oal.ul.pt}}

   \institute{
             Centro de Astronomia e Astrof{\'\i}sica da Universidade de Lisboa,
             Observat\'orio Astron\'omico de Lisboa, Tapada da Ajuda, 1349-018
             Lisboa, Portugal
	 \and    
             Observatoire de Gen\`eve, 
	     51 ch.  des Maillettes, CH--1290 Sauverny, Switzerland
         \and
	     European Southern Observatory, 
	     Casilla 19001, Santiago 19, Chile
	 \and
             Instituto de Astrof{\'\i}sica de Canarias, 
	     E-38200 La Laguna, Tenerife, Spain
 	 \and
	     Laboratoire d'Astrophysique de Marseille, 
	     Traverse du Siphon, 13013 Marseille, France
	 \and
	     Observatoire de Haute Provence, 
	     04870 St Michel l'Observatoire, France	           
	 \and
	     Observatoire de la C\^ote d'Azur,
	     BP 4229, 06304 Nice Cedex 04, France
            }

   \date{Received 24 Nov. 2005; accepted 21 Dec. 2005}

 %\abstract{}{}{}{}{} 
 %5 {} token are mandatory
 
  \abstract
   {During the past years photometric surveys, later complemented
    by follow-up radial-velocity measurements, have revealed the
    presence of several new extra-solar transiting planets, 
    in very short period orbits. Many of the host stars are extremely faint
    (V$\sim$16), making high-precision spectroscopic measurements
    challenging.}
   {We have used the UVES spectrograph (VLT-UT2 telescope) to obtain high resolution 
    spectra of 5 stars hosting transiting planets, namely for OGLE-TR-10, 
    56, 111, 113 and TrES-1. The immediate objective is to derive accurate
    stellar parameters and chemical abundances.}
   {The stellar parameters were derived from an LTE analysis
    of a set of \ion{Fe}{i} and \ion{Fe}{ii} lines. }
   {Complementing the spectroscopic information with photometric transit curves 
    and radial-velocity data from the literature, we have then refined the stellar and 
    planetary radii and masses. The obtained data were also used to study and discuss 
    the relation between the stellar metallicity and orbital period of the planets.}
   {}


\maketitle

\section{Introduction}

%Ten years have passed since the discovery of the first extra-solar planet
%orbiting a solar-type star \citep[][]{Mayor-1995}. Since then, about 160 
%extra-solar giant planets have been discovered orbiting F-G-K stars, showing that
%planets are common in the solar-neighborhood\footnote{For an updated list see 
%table at http://obswww.unige.ch/Exoplanets}. However, the current discoveries
%have also unveiled the presence of giant planets with a huge variety of
%orbital characteristics, thus defying the models of planetary formation and 
%evolution \citep[see e.g.][]{Santos-2005b}. 

Although the majority of known extrasolar planets was discovered using the 
radial-velocity technique\footnote{For a continuously updated list see 
table at http://obswww.unige.ch/Exoplanets}, some major photometric transit 
searches are now delivering important results, giving a new breath 
to the study of exoplanets. Many candidates have been announced by surveys 
like OGLE or TrES \citep[][]{Udalski-2002,Alonso-2004}, and in a few
cases, the planetary nature was confirmed by follow-up radial-velocity
measurements \citep[][]{Konacki-2003,Bouchy-2004,Pont-2004, 
Alonso-2004,Bouchy-2005a,Konacki-2005}. Together with the few transiting 
planets found in the context of radial-velocity surveys 
\citep[][]{Charbonneau-2000,Henry-2000,Sato-2005,Bouchy-2005b}, these
discoveries are now providing information about the physical properties 
of the giant planets themselves (e.g. radius, mean density), and opening 
the possibility to confront the observed properties with those predicted 
by the models.

The new detections have also raised a lot of scientific problems. 
Part of the discovered transiting planets \citep[][]{Konacki-2003,Bouchy-2004} 
have orbital periods of the order of 1-2 earth days. This is in clear contrast 
with the known pile-up of planets with periods above 3-days observed for 
(hot-jupiter) giant planets discovered by the radial-velocity 
technique \citep[see e.g.][]{Santos-2005b}. Whether this is simply the result 
of a detection bias, or if this can be explained by 
some other physical process, is now a matter of debate \citep[e.g.][]{Gaudi-2005}.
On the other hand, among the confirmed transiting planets, 
HD\,209458 \citep[][]{Charbonneau-2000,Henry-2000} has the lowest, and most anomalous 
mean density \citep[][]{Baraffe-2005,Guillot-2002}. Curiously also, a relation between 
the planet mass and the orbital period also seems to exist \citep[][]{Mazeh-2005}. 
The understanding of these issues may give us some important clues about the 
processes of giant planet formation and evolution. 

In order to access these problems we have to be able to derive accurate 
parameters for the planets. But
the derivation of the planetary properties (mass, radius, and as a
consequence their mean density) may depend considerably on the deduced
parameters for the stellar host. For some cases, namely for the
planets discovered in the context of the OGLE survey, the host stars
are extremely faint (V$\sim$16), making difficult the task of obtaining precise
spectroscopic parameters. 

In this paper we present accurate stellar parameters 
and chemical abundances for 5 stars known to be orbited by transiting planets, 
namely for OGLE-TR-10, OGLE-TR-56, OGLE-TR-111 and OGLE-TR-113, as well as 
for the brighter planet-host TrES-1. We use the obtained values to
derive revised stellar and planetary radii and masses, as well as to study the
relation between the presence of planets and the stellar metallicity 
\citep[e.g.][]{Gonzalez-1998,Santos-2001,Santos-2004b} for short orbital 
period planets. In a separate paper we derive and discuss the ages
of stars with short period transiting planets \citep[][]{Melo-2006}.

\section{Observations}
\label{sec:obs}

\begin{table}
\caption{Observations log.}
\label{table:log}
\begin{tabular}{lccccr}
\hline\hline
Star & Mag.$^a$ & T$\mathrm{exp}$ [s] & N(Exp) & Bin & S/N\\
\hline
TrES-1      & 11.8 & 2000 & 1 & 1x1 & 160\\
OGLE-TR-10  & 14.9 & 3000 & 4 & 2x2 & 110\\
OGLE-TR-56  & 15.3 & 3000 & 5 & 2x2 & 100\\
OGLE-TR-111 & 15.5 & 3000 & 6 & 2x2 & 90\\
OGLE-TR-113 & 14.4 & 3000 & 3 & 2x2 & 115\\
\hline
\end{tabular}
\newline
$^a$Magnitudes are expressed in the I-band, except for TrES-1 (in V).
\end{table}

The observations were carried out with the UVES spectrograph at the
VLT-UT2 Kueyen telescope (program ID\,75.C-0185), between April and 
May 2005 (service mode). For all stars, we opted for using a slit
width of 0.9 arcsec, which provides a spectral resolution 
R=$\lambda$/$\Delta\lambda$$\sim$50\,000. The observations
were done using the Dichroic 390+580 mode. The red portion of the
spectra (used in the current paper) covers the wavelength domain 
between 4780 and 6805\AA, with a gap between 5730 and 5835\AA.

For the faint OGLE candidates (I magnitudes between 14.4 and 15.5), 
we have done several exposures of 3000 seconds each. 
For each exposure, the CCD was read in 2x2 bins to reduce the readout noise 
and increase the number of counts in each bin. This procedure does not 
compromise the resolving power, since the sampling of the CCD is still
higher (by a factor of 2) than the instrumental PSF.
For the brighter TrES-1, we kept a 1x1 CCD binning.
The total S/N obtained for each star, as measured directly from small
spectral windows with no clear spectral lines in 
the region near 6500\AA, is listed in Table\,\ref{table:log}.

For the OGLE stars, particular attention was given to the orientation of 
the slit due to the relative crowdedness of the fields. The angle was chosen
in each case using the images available at the OGLE 
website\footnote{http://www.astrouw.edu.pl/$\sim$ftp/ogle/index.html}, 
in order that no other star was present in the UVES slit during the observation.
For some of the targets, like OGLE-TR-10, this can be particularly important.

\begin{table*}
\caption{Stellar parameters and metallicities for known transiting-planet host stars.}
\label{table:parameters}
\begin{tabular}{lcccrccl}
\hline\hline
Star & T$\mathrm{eff}$ [K] & $\log{g}$ (c.g.s.) & $\xi_{\mathrm{t}}$ [km\,s$^{-1}$] & \multicolumn{1}{c}{[Fe/H]} & N(\ion{Fe}{i}, \ion{Fe}{ii}) & $\sigma$(\ion{Fe}{i},\ion{Fe}{ii}) & Source\\
\hline
%\multicolumn{8}{l}{Parameters derived in the current paper:}
\object{TrES-1}      & 5226$\pm$38  & 4.40$\pm$0.10 & 0.90$\pm$0.05 & 0.06$\pm$0.05 & 36,7  & 0.04,0.05 & This paper\\% comparar c/ Sozzetti (2004)
\object{OGLE-TR-10}  & 6075$\pm$86  & 4.54$\pm$0.15 & 1.45$\pm$0.14 & 0.28$\pm$0.10 & 33,11 & 0.08,0.06 & This paper\\% comparar c/ Bouchy (2005)
\object{OGLE-TR-56}  & 6119$\pm$62  & 4.21$\pm$0.19 & 1.48$\pm$0.11 & 0.25$\pm$0.08 & 31,9  & 0.06,0.08 & This paper\\% comparar c/ Bouchy (2005)
\object{OGLE-TR-111} & 5044$\pm$83  & 4.51$\pm$0.36 & 1.14$\pm$0.10 & 0.19$\pm$0.07 & 31,7  & 0.07,0.18 & This paper\\% comparar c/ Pont (2004)
\object{OGLE-TR-113} & 4804$\pm$106 & 4.52$\pm$0.26 & 0.90$\pm$0.18 & 0.15$\pm$0.10 & 30,5  & 0.10,0.09 & This paper\\% comparar c/ Bouchy (2004)
\hline
%\multicolumn{8}{l}{Results compiled from the literature}
\object{OGLE-TR-132} & 6411$\pm$179 & 4.86$\pm$0.14 & 1.46$\pm$0.36 & 	0.43$\pm$0.18 & -- & -- & \citet[][]{Bouchy-2004}\\
\object{HD\,149026}  & 6147$\pm$50  & 4.26$\pm$0.07 & -- 	   & 	0.36$\pm$0.05 & -- & -- & \citet[][]{Sato-2005}\\
\object{HD\,189733}  & 5050$\pm$50  & 4.53$\pm$0.14 & 0.95$\pm$0.07 & $-$0.03$\pm$0.04 & -- & -- & \citet[][]{Bouchy-2005b}\\
\object{HD\,209458}  & 6117$\pm$26  & 4.48$\pm$0.08 & 1.40$\pm$0.06 & 	0.02$\pm$0.03 & -- & -- & \citet[][]{Santos-2004b}\\
\hline
\end{tabular}
\end{table*}

\begin{table*}
\caption{Stellar parameters derived in previous studies for the 5 stars analyzed in the current paper.}
\label{table:compteff}
\begin{tabular}{lcccl}
\hline\hline
Star & T$\mathrm{eff}$ [K] & $\log{g}$ (c.g.s.) & \multicolumn{1}{c}{[Fe/H]} & Source\\
\hline
%\multicolumn{8}{l}{Parameters derived in the current paper:}
TrES-1      & 5214$\pm$23 & 4.52$\pm$0.05 & 0.00$\pm$0.04 & \citet[][]{Laughlin-2005}\\
            & 5250$\pm$75 & 4.6$\pm$0.2   & 0.00$\pm$0.09 & \citet[][]{Sozzetti-2004b}\\
OGLE-TR-10  & 6220$\pm$140& 4.70$\pm$0.40 & 0.39$\pm$0.14 & \citet[][]{Bouchy-2005a}\\
            & 5750$\pm$100& 4.4$^{+0.4}_{-0.9}$ & 0.0$\pm$0.2   & \citet[][]{Konacki-2005}\\
OGLE-TR-56  & 5970$\pm$150& 4.20$\pm$0.38 & 0.17$\pm$0.19 & \citet[][]{Bouchy-2005a}\\
            & $\sim$5950  & --            &  --           & \citet[][]{Konacki-2003}\\
OGLE-TR-111 & 5070$\pm$400& 4.8$\pm$1.0   & 0.12$\pm$0.28 & \citet[][]{Pont-2004}\\
OGLE-TR-113 & 4800$\pm$150& 4.5$^{+0.5}_{-0.8}$ & 0.0$^{+0.1}_{-0.3}$ & \citet[][]{Konacki-2004}\\
            & 4752$\pm$130& 4.50$\pm$0.53 & 0.14$\pm$0.14 & \citet{Bouchy-2004}\\
\hline
\end{tabular}
\end{table*}

\section{Stellar parameters and metallicity}

\begin{figure}
\resizebox{\hsize}{!}{\includegraphics{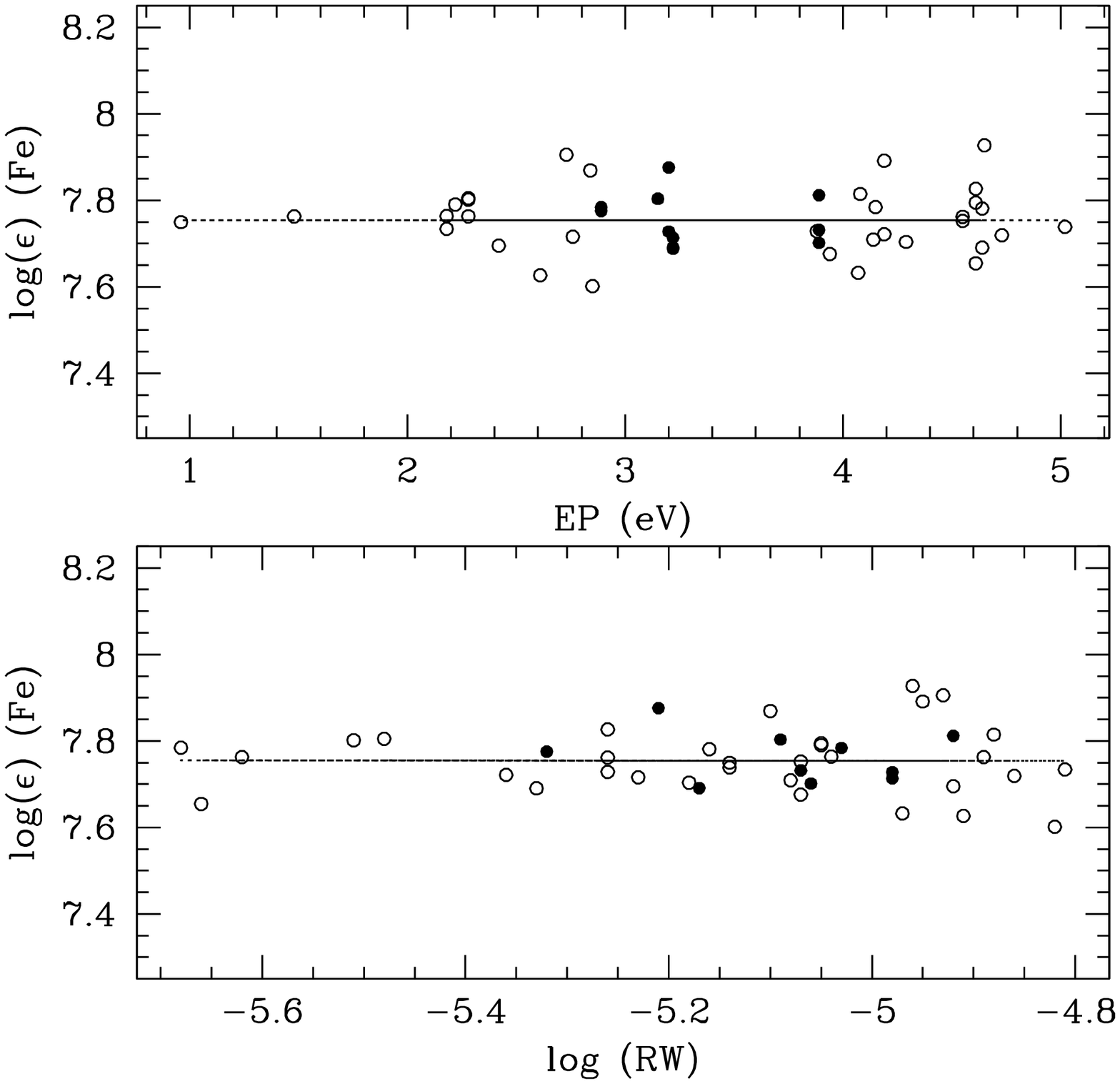}}
\caption{Abundances given by individual \ion{Fe}{i} (open circles) 
and \ion{Fe}{ii} (dots) lines as functions of the lower Excitation Potential 
(top) and the reduced Equivalent Width (bottom) for OGLE-TR-10.
The lines represent least-square fits to the \ion{Fe}{i} line data, and have a null slope. 
To derive the stellar parameters we iterated until the correlation coefficients 
between $\log{\epsilon}$(\ion{Fe}{i}) and EP, and between $\log{\epsilon}$(\ion{Fe}{i}) and the reduced 
equivalent widths were zero, while the abundances derived from the \ion{Fe}{ii} lines were forced 
to be equal to those obtained from \ion{Fe}{i}. For more details on the procedure 
see e.g. \citet[][]{Santos-2004b}.}
\label{fig:feogle10}
\end{figure}

Stellar parameters and chemical abundances were derived 
in LTE using the 2002 version of the code 
MOOG \citep[][]{Sneden-1973}\footnote{http://verdi.as.utexas.edu/moog.html} 
and a grid of Kurucz Atlas plane-parallel model atmospheres \citep[][]{Kurucz-1993}. 
The whole procedure is described in \citet[][and references therein]{Santos-2004b},
and is based on the analysis of 39 \ion{Fe}{i} and 12 \ion{Fe}{ii} weak lines,
imposing excitation and ionization equilibrium (Fig.\,\ref{fig:feogle10}).
As shown in \citet[][]{Santos-2004b}, this methodology gives excellent 
results for the derived stellar parameters of solar-type stars. 

Line Equivalent Widths (EW) for the iron lines were measured using the 
IRAF\footnote{IRAF is distributed by National Optical Astronomy 
Observatories, operated by the Association of Universities for 
Research in Astronomy, Inc.,under contract with the National Science 
Foundation, U.S.A.} ``splot'' and ``bplot'' routines within the 
{\tt echelle} package. 
The final derived stellar parameters and metallicities are presented in 
Table\,\ref{table:parameters}. In this Table we also present the stellar parameters 
for the other stars hosting transiting planets not studied in this paper.
The error bars were derived as 
in \citet[][]{Gonzalez-1996}. In the error determination, the rms ($\sigma$)
around the average abundance given by the \ion{Fe}{i} lines is used 
to compute the final uncertainties in the [Fe/H] abundance, and not 
the $\sigma$/$\sqrt{n}$, with $n$ being the number of lines used. This may 
imply that the error bars are slighly overestimated.

\subsection{Comparison with the literature: the case of OGLE-TR-10}

The values presented in Table\,\ref{table:parameters} for the
5 stars analyzed in the current paper are in 
excellent agreement with previous estimates from the 
literature (see Table\,\ref{table:compteff}). 
The main difference found is on the error bars of the measurements,
that are much smaller in the current study.
The only strong and important disagreement exists for OGLE-TR-10. 

That star has been analyzed before by our team using a S/N$\sim$50 FLAMES/UVES 
spectrum \citep[][]{Bouchy-2005a}, taken with the goal of obtaining 
precision radial-velocities for this star. We derived 
T$_{\mathrm{eff}}$=6220$\pm$140\,k, $\log{g}$=4.70$\pm$0.40 and [Fe/H]=0.39$\pm$0.14, 
values that are close (and compatible within the errors) to the ones derived in the 
current paper. In their paper on OGLE-TR-10, \citet[][]{Konacki-2005} derived a
clearly different set of stellar parameters for this star using a low S/N=44 
spectrum. They have used a spectral fitting procedure to derive the effective temperature,
and obtained (T$_{\mathrm{eff}}$,$\log{g}$,[Fe/H])=(5750\,k,4.4\,dex,0.0\,dex).

The lower T$_{\mathrm{eff}}$ value (by 325\,K) derived 
by \citet[][]{Konacki-2005} can now be reasonably discarded on the basis
of our analysis.
The difference between their value and the one derived in the current paper 
can be due to several effects, including the different spectroscopic analysis 
methods used, the poorer quality of the spectrum taken by \citet[][]{Konacki-2005}, 
or to some other problem related e.g. to the existence of stellar blends (some 
fainter stars exist in the proximity of OGLE-TR-10 -- see Sect.\ref{sec:obs}).
At this moment it is difficult to understand which, if any, of these possibilities
is the cause for the observed discrepancy. Other stars analyzed by both teams
(e.g. OGLE-TR-113) do not show such strong difference in the derived stellar 
parameters.

\begin{figure*}
\resizebox{\hsize}{!}{\includegraphics{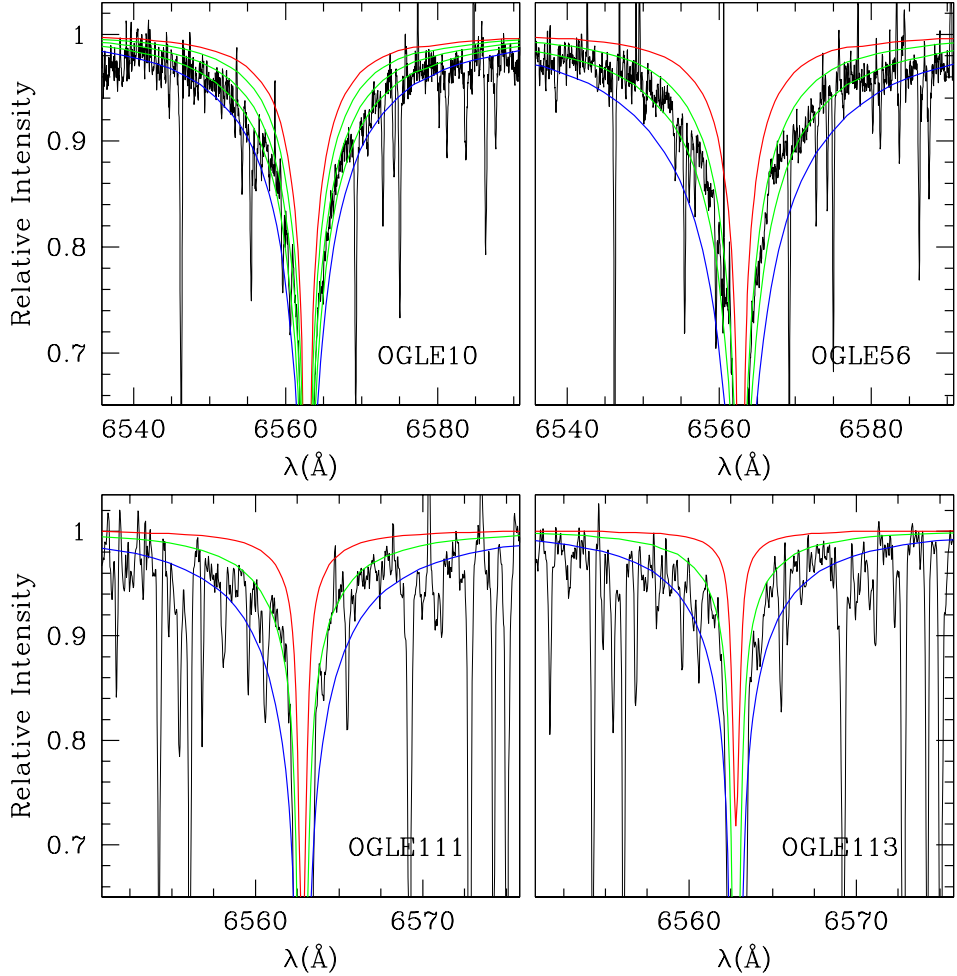}}
\caption{Observed H$_\alpha$ profiles for OGLE-TR-10, 56, 111 and 113, together 
with spectral synthesis for different temperatures, $\log{g}$=4.5 and [M/H]=0.3. 
For OGLE-TR-10, 5 synthesis are shown, for temperatures of (from
top to bottom) 5500, 5750, 6000, 6250 and 6500\,K. For OGLE-TR-111 and 113 three 
synthesis are presented, corresponding to temperatures of
(4500\,K,5000\,K,5500\,K) and (4250\,K,4750\,K,5250\,K), respectively.
For OGLE-TR-56 we present four synthesis, corresponding to temperatures 
of 5500\,K, 6000\,K, 6500\,K and 7000\,K.}
\label{fig:halpha}
\end{figure*}

\subsection{Using H$_{\alpha}$ to check the temperatures}

%Given the relatively low S/N ratio of the individual spectra obtained for 
%the fainter targets in our sample (OGLE candidates), we cannot
%exclude that some systematic errors exist in the data, namely due to
%the data reduction procedure (including the inter-order light subtration).

In order to check the excitation temperatures derived from the previous 
spectroscopic analysis, we have compared the observed H$_\alpha$ line
profiles with synthetic profiles for metal rich dwarfs ([M/H]=0.3 and
$\log{g}$=4.5) computed by R. Kurucz\footnote{These
can be found at http://kurucz.harvard.edu/grids.html.} for different 
temperatures. The results of this comparison can be seen in Fig.\,\ref{fig:halpha}
and show that the H$_\alpha$ temperatures agree with the values 
derived in the previous section for the OGLE stars.

{
It is known that the analysis of H$_\alpha$ profiles can provide accurate 
estimates for the effective temperatures of solar-type dwarfs.
The wings of these lines are very sensitive to T$_{\mathrm{eff}}$ changes, 
while relatively insensitive to surface gravity and metallicity
variations \citep[e.g.][]{Fuhrmann-1993,Barklem-2002}.
However, the fitting of H$_\alpha$ profiles can be subject to 
several sources of error. Given the large width of the line, 
important uncertainties may come from the normalization and continuum determination, 
in particular when dealing with relatively low-S/N echelle spectra 
where one single spectral order spans only a few dozen Angstroms. Furthermore,
the strong line blending present in the spectra of metal-rich dwarfs makes
the H$_\alpha$ line profile determination quite uncertain.
These problems may be at the origin of the very different effective temperature derived
by \citet[][]{Konacki-2005} for OGLE-TR-10. In this paper H$_\alpha$ profiles are thus 
presented only as a confirmation of our Fe-line based temperatures. 
}

\section{The metallicity of stars with short period planets}
\label{sec:metal}

The results presented in Table\,\ref{table:parameters} show that the
stars hosting transiting planets that were studied in the current paper, and
in particular those found by transit surveys like OGLE, follow the general 
trend observed for stars with giant planets: they are metal-rich in comparison 
with the Sun \citep[][]{Gonzalez-1998,Gonzalez-2001,Santos-2001,Santos-2004b,Santos-2005a,Fischer-2005}.

\begin{table*}[t]
\caption{Derived stellar and planetary radii and masses for stars with transiting planets.}
\label{table:planets}
\begin{tabular}{lccccll}
\hline\hline
Star & M$_{\mathrm{star}}$ & R$_{\mathrm{star}}$ & m$_{\mathrm{pl}}$     & r$_{\mathrm{pl}}$ & Reference for  & Reference for \\
     & [M$_{\odot}$]       & [R$_{\odot}$]       & [M${_\mathrm{Jup}}$]  & [R$_{\mathrm{Jup}}$]     & light curve & radial velocity \\
\hline
%\multicolumn{7}{l}{Derived in the current paper:}\\
OGLE-TR-10     & 1.17$\pm$0.04 & 1.14$\pm$0.05 & 0.63$\pm$0.14 & 1.14$\pm$0.09  & \citet[][]{Holman-2005} & \citet[][]{Bouchy-2005a} \\
               &               &               &               &                &                         & \citet[][]{Konacki-2005} \\
OGLE-TR-10     & 1.17$\pm$0.04 & 1.14$\pm$0.05 & 0.63$\pm$0.14 & 1.43$\pm$0.10  & \citet[][]{Udalski-2002c} & \citet[][]{Bouchy-2005a} \\
               &               &               &               &                &                         & \citet[][]{Konacki-2005} \\
OGLE-TR-56     & 1.17$\pm$0.04 & 1.15$\pm$0.06 & 1.24$\pm$0.13 & 1.25$\pm$0.08  & \citet[][]{Udalski-2002c}   & \citet[][]{Konacki-2003}\\
               &               &               &               &                &                         & \citet[][]{Torres-2004} \\
               &               &               &               &                &                         & \citet[][]{Bouchy-2005a} \\
OGLE-TR-111    & 0.81$\pm$0.02 & 0.83$\pm$0.02 & 0.52$\pm$0.13 & 0.97$\pm$0.06  & \citet[][]{Udalski-2002b}   & \citet[][]{Pont-2004} \\
\hline
%\multicolumn{7}{l}{From the literature:}\\
%TrES-1$^a$         & 0.87$\pm$0.03 & 0.83$\pm$0.03   &0.729$\pm$0.036& 1.08$\pm$0.05  & -- &  -- \\
%OGLE-TR-113$^b$    & 0.77$\pm$0.06 & 0.765$\pm$0.025 & 1.35$\pm$0.22 & 1.08$^{+0.07}_{-0.05}$ & -- & --\\
%OGLE-TR-132$^c$    & 1.35$\pm$0.06 & 1.43$\pm$0.10   & 1.19$\pm$0.13 & 1.13$\pm$0.08  & -- &  --  \\
%HD\,149026$^d$    & 1.3$\pm$0.1   & 1.45$\pm$0.1    & 0.36$\pm$0.03 & 0.725$\pm$0.05 & -- &  -- \\
%HD\,189733$^e$     & 0.82$\pm$0.03 & 0.76$\pm$0.01   & 1.15$\pm$0.04 & 1.26$\pm$0.03  & -- & -- \\
%HD\,209458$^f$     & 1.11$\pm$0.16 & 1.12$\pm$0.05   & 0.69$\pm$0.02 & 1.32$\pm$0.05 & -- &  -- \\
%\hline
\end{tabular}
%\newline 
%The parameters were taken from: $^a$\citet[][]{Laughlin-2005}; $^b$\citet[][]{Bouchy-2004}; $^c$\citet[][]{Moutou-2004}; $^d$\citet[][]{Sato-2005}; $^e$\citet[][]{Bouchy-2005b}; $^f$\citet[][]{Laughlin-2005} and \citet[][]{Brown-2001}
\end{table*}

The stars hosting the three shortest period planets listed in the 
Table (OGLE-TR-56 and 113, analyzed in the current paper, 
and OGLE-TR-132, with a less reliable [Fe/H] value, all with orbital periods 
shorter than 2-days) also follow this trend. But although they do not present 
metallicities significantly above the rest of the sample, a small 
``excess'' cannot be excluded. For instance, the lower metallicity amid
the stars with planets in orbital periods below 2 days is $+$0.15 (OGLE-TR-113), while
this value decreases to $-$0.03 (HD\,189733) in the orbital period range from 2-3 days.

{
In order to check if some metallicity trend exists among
the stars with short period planets, we have used the spectroscopic data from Table\,\ref{table:parameters}, 
together with the metallicity values listed in \citet[][]{Santos-2004b} and \citet[][]{Santos-2005a}, as well as the orbital periods listed in the Geneva web page\footnote{http://obswww.unige.ch/Exoplanets}, 
to derive the average [Fe/H] of stars with planets in bins of orbital period. 
The results show that stars with planets having orbital periods below 2 days (OGLE-TR-56, 
113 and 132) have an average metallicity of $+$0.28 ($\sigma$=0.14\,dex, 3 stars -- excluding OGLE-TR-132 this value would decrease to $+$0.20). This average metallicity decreases
to $+$0.23 ($\sigma$=0.14) in the period interval between 2 and 3 days (7 stars), 
to $+$0.21 ($\sigma$=0.11) for periods between 3 and 5 days (17 stars),
and to $+$0.19 ($\sigma$=0.22) for periods between 5 and 10 days (5 stars)\footnote{The
pile-up of planets in the orbital period regime around 3-5 days has been
thoroughtly discussed in the literature \citep[e.g.][]{Udry-2003,Gaudi-2005}.}.
Note that here we have excluded the M-dwarf planet-hosts, as their metallicities 
are less reliable.
}

{
A word of caution should be mentioned to say that particular care must be taken 
when looking at these results. First of all, the number of stars is statistically
small. Furthermore, a significant part of them comes from different galactic regions
and populations. The OGLE objects are faint and
distant stars (at a few Kpc from the Sun) seen in the direction of the Galactic bulge 
(OGLE-TR-10 and 56) and Carina arm (OGLE-TR-111, 113 and 132). Given
the radial metallicity gradient in the Galaxy \citep[][]{Nordstrom-2004}, they may belong to different
populations when compared to the solar neighborhood field stars studied 
in radial-velocity surveys \citep[see discussion in][]{Sozzetti-2004}. For instance, 
the average metallicity of the OGLE stars (that span a large range of effective 
temperatures and orbited by planets with periods from $\sim$1.2 to $\sim$4 days) is
$+$0.26\,dex, clearly above the average value for the other 4 
stars in Table\,\ref{table:parameters} ($+$0.10).
}

{
It has been suggested that stars orbited by short period planets 
(hot- and very-hot-jupiters) may have an average metallicity above the one found 
for other planet-hosts \citep[][]{Gonzalez-1998,Queloz-2000,Santos-2003,Sozzetti-2004,Fischer-2005}.
As we have seen above, whether this trend is real or even intensified 
for the very-hot-jupiters is not clear, but if this were the case, 
several possible explanations exist. 
}

{
The possibility that planet 
migration is somewhat controlled by the dust content of the disk seems to 
be reasonably ruled out by current models \citep[][]{Livio-2003}, that generaly
predict that migration rates are not strongly affected by the dust content of the gas 
disk. On the other hand, although migration induced by interactions with
the disk of planetesimals is possible, it seems to require a very high value for the 
mass of the planetesimal disk \citep[e.g.][]{Murray-1998}.
} 
 
{
Another possibility is that a more significant migration could increase
planetesimal scattering into the star and therefore pollute the stellar outer 
layers. This would lead to an increase of the observed stellar 
iron-to-hydrogen ratio \citep[e.g.][]{Murray-2001}. The lack of a general
correlation between the depth of the convective zone (or alternatively the
effective temperature) and the metallicity of the star implies that this possibility is unlikely \citep[][]{Pinsonneault-2001,Santos-2003,Fischer-2005}, although probably
not completely excluded \citep[][]{Vauclair-2004}.
We should add that we do find a weak correlation between effective temperature 
and stellar metallicity in stars with short period planets (P$<$10 days). However, not only 
this correlation is not significant but also it may simply
reflect biases in the samples like the cut in B-V of the radial-velocity planet-search 
samples \citep[e.g.][]{Santos-2004b} and/or the galactic chemical evolution gradient \citep[][]{Edvardsson-1993a}.
}

\begin{figure}[h]
\resizebox{\hsize}{!}{\includegraphics{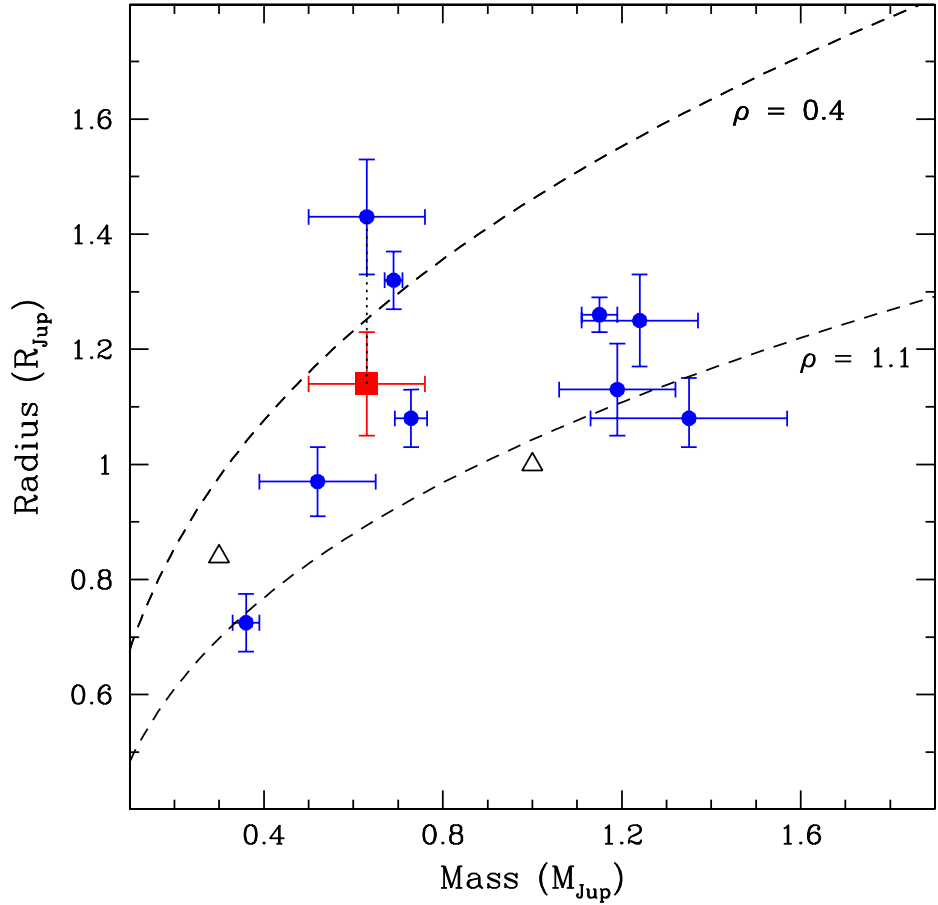}}
\caption{Mass-radius diagram for the known transiting planets. For OGLE-TR-10b we
show both values for the radius listed in Table\,\ref{table:planets}, the filled square
denoting the values derived using the unpublished photometry by \citet[][]{Holman-2005}.
A dotted line connects the two different points of OGLE-TR-10b.
Two iso-density curves are also shown for mean densities of 0.4 and 1.1\,g\,cm$^{-3}$.
The open triangles denote the positions of Jupiter and Saturn. The data were taken
from Table\,\ref{table:planets} for OGLE-TR-10b, 56b and 111b, and 
from \citet[][for TrES-1b]{Laughlin-2005}, \citet[][OGLE-TR-113b]{Bouchy-2004}, 
\citet[][OGLE-TR-132b]{Moutou-2004}, \citet[][HD\,149026b]{Sato-2005}, 
\citet[][HD\,189733b]{Bouchy-2005b}, \citet[][]{Laughlin-2005} 
and \citet[][]{Brown-2001} (HD\,209458b).}
\label{fig:massradius}
\end{figure}

An alternative explanation for the slight metallicity-period trend could have to do with
planetary internal structure arguments. If a planet forming in a higher
metallicity disk acquires a more massive core
or has a higher metal-to-hydrogen ratio \citep[e.g.][]{Pollack-1996,Ida-2004b,Guillot-2006}, 
its higher density will help its survival against 
evaporation \citep[][]{Baraffe-2004,Etangs-2004}. 
We note that at least in one case evidence was found for a hot-jupiter having a core mass close
to 70\,M$_\oplus$, and orbiting a very metal-rich star \citep[][]{Sato-2005}.
This idea could, in principle, also be able to explain 
any possible metallicity-effective temperature correlation, if one considers
that only the most ``metal-rich'' planets would survive close to a higher temperature 
dwarf emitting a higher UV-flux.

Finally, such a correlation may arise indirectly from the combination
of a faster growth of planets and their migration. This occurs in core-accretion 
models in which the loss of the gas disk occurs on timescales similar to 
the growth of the planets. In that case, a more rapid growth in a metal-rich 
environement will imply a more pronounced migration \citep[see e.g.][]{Lecar-2003,Ida-2004b,Alibert-2005,Guillot-2006b}.

\section{Revised planetary masses and radii}
\label{sec:planet}

Using the stellar parameters presented in Table\,\ref{table:parameters}, 
we have re-calculated the planetary and stellar radii and masses following to the
method described in \citet[][]{Bouchy-2005a}. The constraints from the spectroscopic
parameters and the photometric transit shape are combined by $\chi^2$
minimisation to derive the mass and radius of the host star under the
assumption that it is a main-sequence star of any age described by 
the \citet[][]{Girardi-2000} stellar evolution models. The radii of the planets are obtained
from the radius ratio given by the light curve, and the planet's mass is
derived from the mass function given by the radial velocity orbit. 

In this procedure we adopted the latest and best observations known to us for the 
light curves and radial velocity orbits (see Table\,\ref{table:planets} for references).
For OGLE-TR-111, we also used the new ephemerids found by the OGLE team from recent 
observations: $T_{tr}$=2452330.46228, $P$=4.014442 days (A. Udalski, priv. comm.). 
For OGLE-TR-10 we have used both the original photometry from the
OGLE campaign and the new results from \citet[][]{Holman-2005}.

Table\,\ref{table:planets} gives the masses and radii values for host stars and 
planets derived from our analysis. We point out that the error bars do not include 
the systematic uncertainties on the host star properties due to the uncertainties 
on stellar evolution models, bolometric corrections, and atmosphere models. 
Given the similarity of stellar parameters derived in this paper
with those found in the literature for OGLE-TR-113 and TrES-1 we have not derived
revised parameters for these two stars.

In Fig.\,\ref{fig:massradius} we plot a mass-radius diagram for the known
transiting planets. Without a detailed evolution model accounting for the planet's mass, 
age and stellar insolation, one cannot directly infer the composition of the planets 
from such a diagram \citep[e.g.][]{Guillot-2005}. However, as also previously 
noted in different works \citep[e.g.][]{Baraffe-2005,Laughlin-2005}, 
both HD\,209458b and OGLE-TR-10b 
\citep[considering the original result by][]{Bouchy-2005a} appear 
to have anomalous mean densities. For this latter case the use
of the recent photometry by \citet[][]{Holman-2005} together with
the stellar parameters derived in the current paper seem to put this 
planet in a rather normal position of this diagram. 

Specific evolution models (Guillot et al., in prep.) indicate that, 
although it appears relatively ``normal'' in Fig.\,\ref{fig:massradius}, 
HD\,189733b is also anomalously large. This is due to its relatively 
large mass and low insolation that yield evolution models to predict 
a smaller radius.

\section{Concluding remarks}

We have used high resolution and S/N spectra taken with the UVES spectrograph (VLT-UT2)
to derive accurate stellar parameters and metallicities for 5 stars known
to be transited by giant planets. The resulting parameters were used to study both the
relation between the stellar metallicity and the existence of giant planets
in short period orbits, and to derive revised masses and radii for the transiting 
planets.

The results show that the stars with transiting planets studied in the
current paper follow the general trend of high metallicities found for
the majority of the planet hosts \citep[][]{Gonzalez-1998,Santos-2001}.
A weak, but not significant correlation is found for the average metallicity 
as a function of the orbital period among the short period planets. The addition
of more data is needed to confirm/deny this result.

\begin{acknowledgements}
   We would like to thank our referee, A. Sozzetti, for the
   useful comments and suggestions.
   Support from Funda\c{c}\~ao para a Ci\^encia e a Tecnologia (Portugal)
   to N.C.S. in the form of a scholarship (reference SFRH/BPD/8116/2002)
   and a grant (reference POCI/CTE-AST/56453/2004) is gratefully
   acknowledged.
\end{acknowledgements}

\bibliographystyle{aa}
\bibliography{santos_bibliography}

\end{document}